# Highly Conductive 3D Nano-Carbon: Stacked Multilayer Graphene System with Interlayer Decoupling

Tianhua Yu[†], Changdong Kim[†], and Bin Yu*[,†]

[†] College of Nanoscale Science and Engineering, State University of New York, Albany, NY 12203

* Corresponding Author.  byu@uamail.albany.edu

**ABSTRACT**   We investigate electrical conduction and breakdown behavior of 3D nano-carbon – stacked multilayer graphene (s-MLG) system with complete interlayer decoupling. The s-MLG is prepared by transferring and stacking large-area CVD-grown graphene monolayers, followed by wire patterning and plasma etching. Raman spectroscopy was used to confirm the layer number. The D-band peak indicates low defect level in the samples. Electrical current stressing induced doping is performed to shift the charge-neutrality Dirac point and decrease the graphene/metal contact resistance, improving the overall electrical conduction. Breakdown experiments show the current-carrying capacity of s-MLG is largely enhanced as compared with that of monolayer graphene.

Graphene, a 2D allotrope of carbon,[1] is a superb conductor that is potentially useful in a broad spectrum of applications including nano-electrodes for emerging energy conversion devices [2] and on-chip interconnects for information-processing systems. [3] Exceptional properties of graphene have been reported, including ultra-high carrier mobility (up to 200,000 $cm^2$/V-s at room temperature) and thermal conductivity (4800 W/m-K, ~10 times of Cu).[4,5] The excellent electrical, thermal, and mechanical properties, as well as the immunity to electromigration makes graphene an ideal material candidate for high-performance interconnects.[6-8] Monolayer graphene nanoribbon (GNR) is expected to outperform Cu at ultra-scaled width dimensions.[9] Although the maximum current-carrying density up to $10^8$ $A/cm^2$ (~100 times of Cu) was reported, current density of monolayer graphene is limited by its 2D atomically-thin geometry, sensitivity to ambient, poor noise tolerance, and conductance degradation due to edge-related effects.[10,11] 3D nano-carbon, or vertically stacked multilayer graphene (s-MLG), can potentially be employed to address this challenge, yielding an ultra-conductive material system. Graphene multilayer has received significant amount of theoretical and experimental attentions in recent years[12]. However, in the case of *Bernal stacking* (in which graphene layers have an *ABAB*-ordered lattice arrangement), the composing graphene monolayers are electronically coupled and form graphite which lacks the appealing electronic properties of individual 2D graphene sheets. It is therefore ideal to have *electronically decoupled* layers in 3D stacked multilayer system. In this letter, we investigate electrical conduction and breakdown characteristics of doped 3D stacked multilayer graphene (layer number (*n*) up to 5) with complete interlayer decoupling.

Figure 1 shows the fabrication process flow for the s-MLG system. First, graphene monolayer was grown on 25 μm thick Cu foil in a quartz-tube furnace system using low-pressure CVD (chemical vapor deposition) method involving methane (as precursor) and hydrogen gases by taking advantage of the low solubility of carbon in Cu.[13] After the growth, one side of graphene sheet was covered with polymethyl methacrylate (PMMA), Cu was then etched away by an aqueous solution of iron nitrate (0.05 g/mL) over a time period of 12 hours. The PMMA/graphene stack was washed with de-ionized water and placed on the target substrate (SiO$_2$-coated Si wafer) and dried.[14] By repeating the same CVD growth and layer-transfer process, multilayer graphene was stacked in an uncorrelated (or decoupled) way on the new target substrate. The sample quality was evaluated by optical microscopy and micro-Raman spectroscopy. Electron beam lithography (Vistec VB300 system) was used to pattern graphene wire as well as to define metal contacts.[15] Oxygen plasma etching was used to pattern the s-MLG wires with different sizes (down to sub-100nm width). A metal liftoff process was used to form the probing contacts made of Ti/Au stack. Agilent B1500A semiconductor parameter analyzer was used for electrical characterization. Testing was performed with standard lock-in methods. All measurements in this work were performed at room temperature.

Raman spectral analysis is particularly useful in determining the number of graphene layers, strain level, and defect concentration.[16] Figure 2a shows the Raman spectra of different s-MLG samples with the graphene layer number *n* varying from 1 to 5, as compared with the reference: KISH graphite. The intensity of D-band peak at about 1350 cm$^{-1}$ is used as a signature of defects which allows one to determine the quality of a sample.[17] The insignificant D-band peak in all our tested samples suggests that defects are minimal in the prepared graphene multilayers (even for *n* = 5). The primary observed defects are wrinkles which were generated during the CVD growth as a result of the difference in thermal expansion coefficients between graphene and copper. The measured Raman spectra show sharp, symmetrical G-band and 2D-band peaks for the transferred graphene monolayer, indicating high quality of samples. The Raman spectra of the stacking multilayer graphene show considerable difference as compared with those micromechanically cleaved (or exfoliated) graphene sheets obtained from highly oriented pyrolytic graphite (HOPG) crystals. The 2D peak does not exhibit obvious shoulder feature as in *ABAB*-order graphene multilayer. It is much broader and almost resembles a single-peak as noticed in a single layer of graphene, possibly due to the reduced interlayer coupling between the transferred graphene monolayers. Upon analyzing the features of both the G-band and 2D-band peaks, we could find two key signatures for indentifying the physical thickness (or layer number) of the s-MLG samples: the G–to–2D intensity ratio and the 2D peak location. Both are increased with increasing layer number, as shown in Figure 2b and 2c. Similar results were reported on exfoliated multilayer graphene.[18]

Doping is necessary to enhance electrical conduction in 3D s-MLG system. The shift in Fermi energy that naturally occurs because of the charge trapped at the graphene-substrate interface is limited to the first few graphene monolayers at the bottom of multilayer stack due to the screening effect.[19] Unlike copper or other metals, doping can be induced in graphene system to further decrease the electrical resistance.[20] A variety of doping strategies have been previously employed in exfoliated graphene.[21] Among different approaches, electrical-stress-induced doping has been demonstrated to be clean and time-efficient, without the involvement of additional gate or absorbents.[22] As shown in Figure 3a, if the electrical stressing voltage (applied on one end of the wire, while keeping the other end grounded) is raised up to 40V, the Dirac point of s-MLG (measured from the resistance-voltage or R-V curve with backgate sweeping) shifts towards right, which implies holes injection into the graphene. Although the doping mechanism of high-level electrical stressing was not fully understood yet, electrostatic modifying of the substrate surface (involving the filling/releasing of deep-level charge traps) has been considered as the possible reason.[23] Due to the shift of Dirac point, electrical resistance of s-MLG wire is greatly decreased at zero-bias back-gate voltage (~3 times lower).

To better interpret the mechanism of total resistance decrease in the s-MLG system, the transfer length method (TLM) is employed to probe the resistance between graphene and metal contact (Ti/Au).[24] As in Figure 3b, it is observed that both graphene/metal contact resistance and graphene wire resistance keep decreasing with increasing level of stressing voltage. It is believed that the conduction improvement in graphene wire is due to doping effect and absorbent removal from graphene surface as a result of Joule heating, while the improvement in contact resistance is likely attributed to thermal annealing effect. Therefore, electrical annealing could be an efficient way to improve the overall electrical conduction of the s-MLG system.

The maximum current-carrying density is crucial for a conductor system[10]. The maximum current density in Cu is reported to be approximately $10^6$ A/cm$^2$, a severe restriction for nanometer-sized wiring systems.[25] Current-stressing is applied on the s-MLG samples by gradually increasing the voltage bias until permanent breakdown occurs. As seen in Figure 4a, when the s-MLG sample (n = 5) is tested under the electrical stressing, a physical break is clearly seen in the middle of the graphene wire. Figure 4b shows the I-V characteristics during the stressing test. As current density is increased to the threshold level (at which breakdown occurs), an abrupt drop in current is observed. Given the s-MLG width of 2 μm and the flake thickness of 1.6 nm (assuming 0.35 nm for monolayer graphene), the threshold current density at hard breakdown is calculated to be over $1.1 \times 10^9$ A/cm$^2$, which is more than 5 times that of a graphene monolayer. The improved current-carrying capability is attributed to the addition of more graphene sheets as conduction layer.

In summary, the electrical conduction and breakdown behavior of 3D vertically stacked multilayer graphene system is investigated. The s-MLG samples were prepared by physical stacking of transferred large-area graphene monolayers (assembled via the CVD-based growth process). Low-level of defects in transferred graphene monolayer is confirmed by Raman spectra. Electrical stressing induced doping and thermal annealing are found to be effective to considerably improve the electrical conduction of both s-MLG and s-MLG/metal contact. Experiment on 3D s-MLG system shows largely improved physical breakdown threshold and maximum current-carrying density over that of 2D monolayer graphene system.

**Acknowledgment.** The research was partially supported by National Science Foundation (NSF) grants (ECCS-1002228 and ECCS-1028267) and IBM Faculty Award.

**REFERENCES:**

1. A. K. Geim and K. S. Novoselov, Nature Mater. **6**, 183 (2007).

2. X. Wang, L. Zhi and K. Muller, Nano Lett. **8**, 323 (2008).

3. D. Sarkar, C. Xu, H. Li and K, Banerjee, IEEE Trans, Elec. Dev. **58**, 843 (2011).

4. X. Du, I. Skachko, A. Barker, and E. Y. Andrei, Nat. Nano. **3**, 491 (2008).

5. A. A. Balandin, S. Ghosh, W. Bao, I. Calizo, D. Teweldebrhan, F. Miao, and C. N. Lau, Nano Lett. **8**, 902 (2008).

6. T. Saito, T. Yamada, D. Fabris, H. Kitsuki, P. Wilhite, M. Suzuki, and C. Yang, Appl. Phys. Lett. **93**, 102108 (2008).

7. P. G. Collins, M. Hersam, M. Arnold, R. Martle, and Ph. Avouris, Phys. Rev. Lett. **86**, 3128 (2001).

8. H. Kitsuki, T. Yamada, D. Fabris, J. Jameson, P. Wilhite, M. Suzuki, and C. Yang, Appl. Phys. Lett. **92**, 173110 (2008).

9. A. Naeemi and J. D. Meindl, IEEE Electron Dev. Lett., **28**, 428 (2007).

10. T. Yu, E. Lee, B. Briggs, B. Nagabhirava, B. Yu, IEEE Electron Dev. Lett. **31**, 1155 (2010).

11. R. Murali, Y. Yang, K. Brenner, T. Beck, and J. D. Meindl, Appl. Phys. Lett. **94**, 3114 (2009).

12. Y. Sui and J. Appenzeller, Nano Lett. **9**, 2973 (2009).

13. X. Li, W. Cai, J. An, S. Kim, J. Nah, D. Yang, R. Piner, A. Velamakanni, I. Jung, E. Tutuc, S. Banerjee, L. Colombo, R. Ruoff, Science **324,** 1312 (2009).

14. X. Li, Y. Zhu, W. Cai, M. Borysiak, B. Han, D. Chen, R. Piner, L. Colombo, and R. Ruoff, Nano Lett., **9**, 4359 (2009).

15. R. Murali, K. Brenner, Y. Yang, T. Beck, and J. D. Meindl, IEEE Electron Dev. Lett. **30**, 611 (2009).

16. A. Ferrari, J. Meyer, V. Scardaci, C. Casiraghi, M. Lazzeri, F. Mauri, S. Piscanec, D. Jiang, K. Novoselov, S. Roth, and A. Geim, Phys. Rev. Lett. **97**, 187401 (2006).

17. A. Das, B. Chakraborty and A. Sood, Bull. Mater. Sci. **31**, 579 (2008).

18. D. Graf, F. Molitor, K. Ensslin, C. Stampfer, A. Jungen, C. Hierold, and L. Wirtz, Nano Lett. **7,** 238 (2007).

19. H. Wang, Y. Wu, Z. Ni, and Z. Shen, Appl. Phys. Lett. **92**, 053504 (2008).

20. F. Traversi, V. Russo, and R. Sordan, Appl. Phys. Lett. **94**, 223312 (2009).

[21] K. Brenner and R. Murali, Appl. Phys. Lett., **96**, 063104 (2010).

[22] J. Moser, A. Barreiro, and A. Bachtold, Appl. Phys. Lett. **91**, 163513 (2007).

[23] H. Chiu, V. Perebeinos, Y. Lin, and P. Avouris, Nano Lett., **10,** 4634 (2010).

[24] S. Russo , M.F. Craciun, M. Yamamoto, A. Morpurgo, S. Tarucha, Physica E **42**, 677 (2010).

[25] P. Wang and R. G. Filippi, Appl. Phys. Lett. **78**, 3578 (2001)

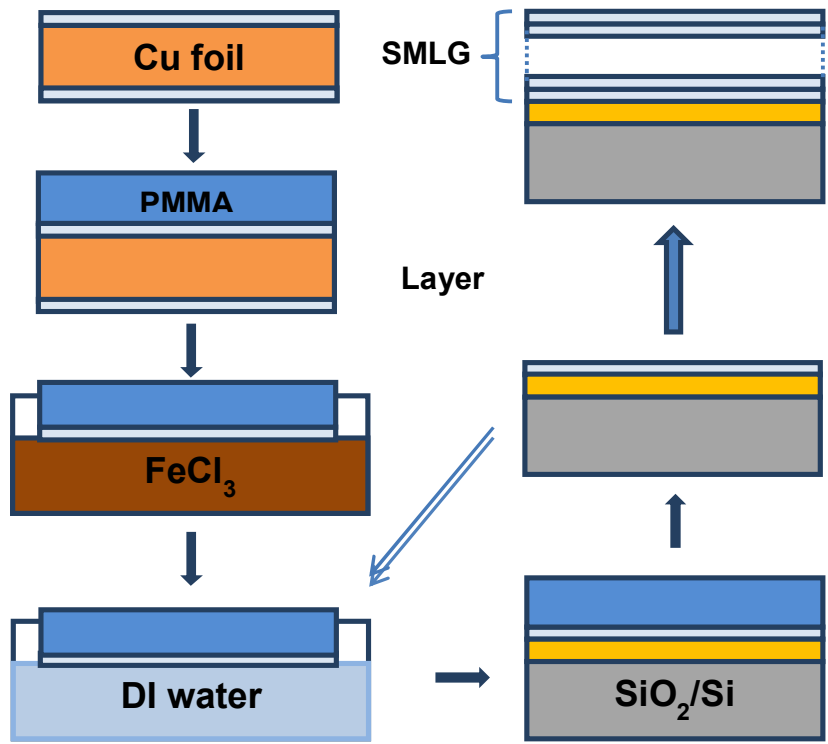

**Figure 1.** Schematic showing the preparation process of 3D decoupled graphene multilayer via repeatedly stacking the transferred graphene monolayers (via CVD-based growth).

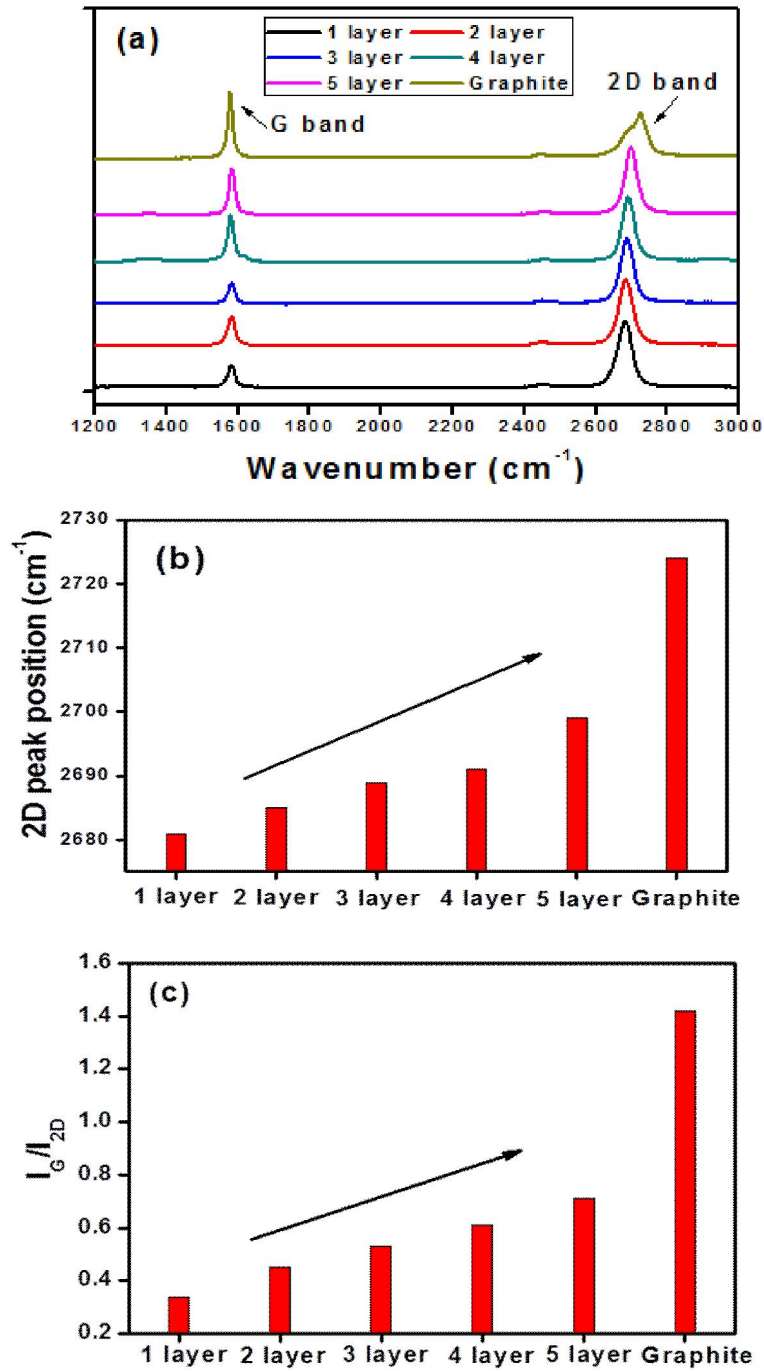

**Figure 2.** Key signatures acquired from Raman spectroscopy for stacking multilayer graphene on $SiO_2$/Si substrate. **(a)** n = 1 to 5 layers of s-MLG (with KISH graphite as the reference). **(b)** Plot of the ratio of the G-to-2D peak intensities vs. number of stacked graphene layer. **(c)** 2D peak location vs. number of stacked graphene layer.

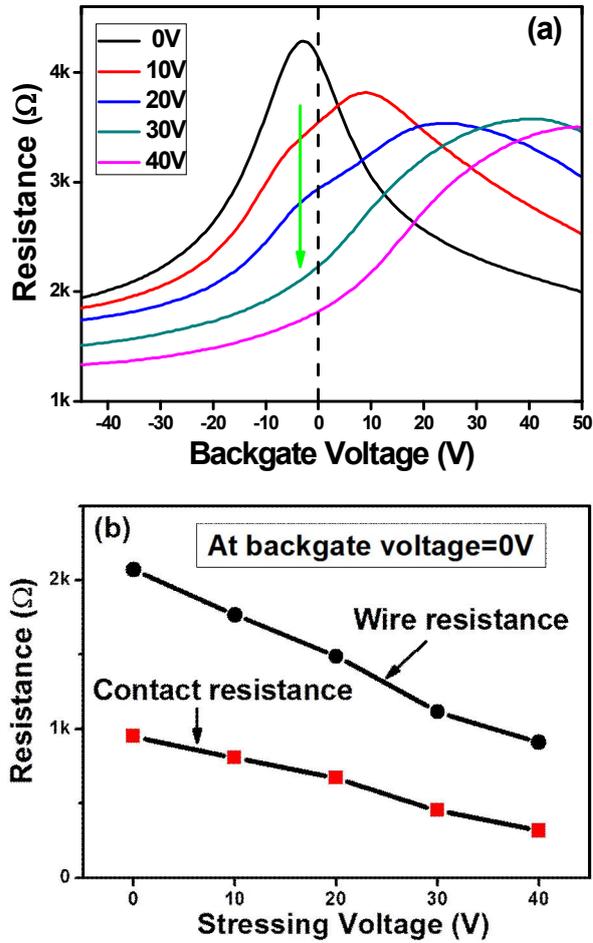

**Figure 3.** Electrical stressing induced effect on electrical conduction of 3D s-MLG samples: **(a)** Measured total resistance (graphene wire and graphene/metal contact) vs. back-gate sweeping voltage with different stressing voltages. **(b)** Contact resistance and wire resistance vs. different stressing voltages.

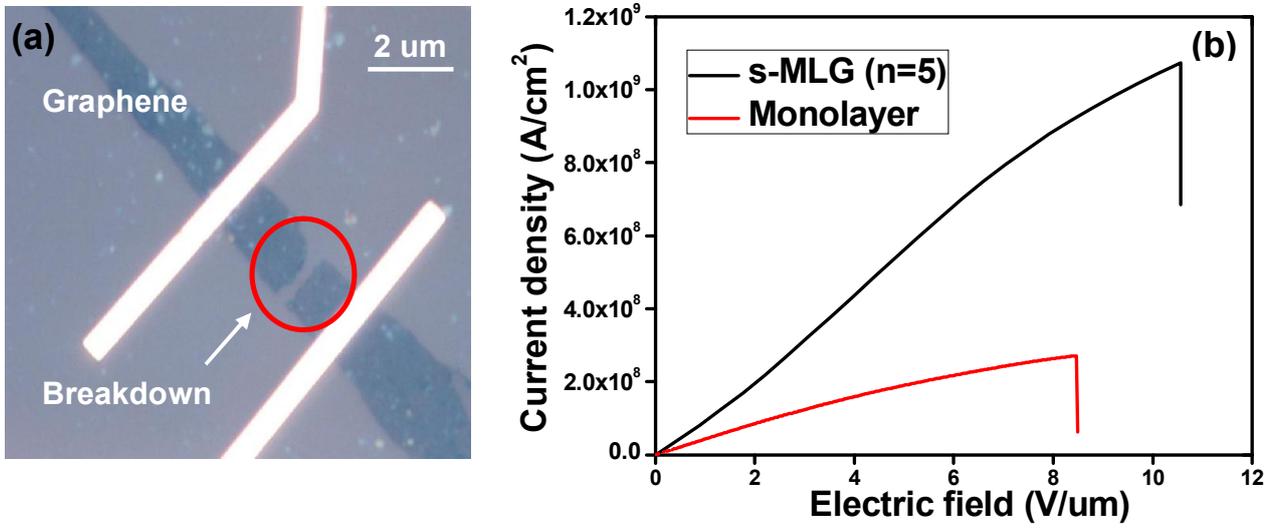

**Figure 4.** Breakdown test of both s-MLG (n = 5) and CVD-grown/transferred graphene monolayer: **(a)** Optical image of the failure spot in the s-MLG. **(b)** Measured I-V characteristics under electrical stressing.